\documentclass{aa}
\usepackage{graphicx}
\usepackage{txfonts}
\usepackage{natbib}
\usepackage{xspace}
\usepackage{dcolumn}

\newcommand{\msun}{\mbox{${\rm M}_{\odot}$}\xspace}
\newcolumntype{d}{D{.}{.}{-1}}
\newcolumntype{m}{D{+}{\mbox{$\pm$}}{-1}}
\def\rev#1{#1}
\def\2rev#1{#1}

\begin{document}
   \title{Binaries discovered by the SPY project}

   \subtitle{IV. Five single-lined DA double white dwarfs\thanks{Based
   on observations made with the INT and WHT operated on the island of
   La Palma by the Isaac Newton Group in the Spanish Observatorio de
   Roque de los Muchachos of the Instituto de Astroph\'isica de
   Canarias}\fnmsep\thanks{Based on observations at the Paranal and La
   Silla Observatories of the European Southern Observatory for
   program Nos. 165.H-0588, 167.D-0407, 68.D-0483, 69.D-0534,
   70.D-0334, 71.D-0383}\fnmsep\thanks{Based on observations collected
   at the German-Spanish Astronomical Center (DSAZ), Calar Alto,
   operated by the Max-Planck-Institute f\"ur Astronomie jointly with
   the Spanish National Commission for Astronomy}}

   \author{G. Nelemans
          \inst{1,2}
          \and
          R. Napiwotzki\inst{3}
          \and
          C. Karl\inst{4}
          \and
          T. R. Marsh\inst{5}
          \and
          B. Voss\inst{6}
          \and
          G. Roelofs\inst{1}
          \and
          R. G. Izzard\inst{2,7}
          \and
          M. Montgomery\inst{2}
          \and
          T. Reerink\inst{8}
          \and 
          N. Christlieb\inst{9}
          \and
          D. Reimers\inst{9}
          }

   \offprints{G. Nelemans}

   \institute{Department of Astrophysics, Radboud University Nijmegen, PO Box 9010 6500 GL, Nijmegen, The Netherlands\\
     \email{nelemans@astro.ru.nl}
     \and
     Institute of Astronomy, University of Cambridge, Madingley Road, Cambridge CB3 0HA, UK
     \and
Centre for Astrophysics Research, University of Hertfordshire, Hatfield, AL10 9AB, UK
     \and
     Remeis-Sternwarte, Universit\"at Erlangen-N\"urnberg, Sternwarte str. 7, 96049 Bamberg, Germany
     \and
     Department of Physics, University of Warwick, Coventry CV4 7AL, UK
     \and
     Institut f\"ur Theoretische Physik und Astrophysik, University of Kiel, 24098 Kiel, Germany
     \and
     The Carolune Institute for Quality Astronomy, www.ciqua.org, UK
     \and
     Astronomical Institute ``Anton Pannekoek'', University of Amsterdam, Kruislaan 403, 1098 SJ Amsterdam, The Netherlands
     \and
     Hamburger Sternwarte, Universit\"at Hamburg, Gojensbergweg 112, 21029 Hamburg, Germany
        }

   \date{Received \today; accepted }

   \abstract{ 

We present results from our ongoing follow-up observations of double
 white dwarf binaries detected in the ESO SN Ia Progenitor SurveY
 (SPY). We discuss our observing strategy and data analysis and
 present the orbital solutions of five close double white dwarf
 binaries: HE0320$-$1917, HE1511$-$0448, WD0326$-$273, WD1013$-$010
 and WD1210+140. Their periods range from 0.44 to 3.22 days. In none
 of these systems we find any spectral lines originating from the
 companion. This rules out main sequence companions and indicates that
 the companion white dwarfs are significantly older and cooler than
 the bright component. Infrared photometry \rev{suggests} the presence
 of a cool, helium-rich white dwarf companion in the binary
 WD\,0326$-$273.  We \rev{briefly} discuss the consequences of our
 findings for our understanding of the formation and evolution of
 double white dwarfs.

   }

   \maketitle
%

\section{Introduction}

\begin{table*}
\caption{Overview of the follow-up observations, giving dates,
  telescopes, instruments and observers. For details of the set-ups of
  the different instruments, see text.}
\label{tab:obs}
\begin{center}
\begin{tabular}{lllll}\hline \hline
Date             & Observatory & Telescope & Instrument & Observer \\
\hline
2001-03-10 to 13 & Calar Alto  & 3.5m      & TWIN       & R. Napiwotzki,
E.-M. Pauli \\
2001-07-06 to 15 & Calar Alto  & 3.5m      & TWIN       & R. Napiwotzki,
C. Karl \\
2001-10-11 to 12 & Paranal     & UT2       & UVES       & R. Napiwotzki
\\
2001-10-26 to 01 & La Palma & INT & IDS & G. Nelemans, T. Reerink \\
2002-02-22 to 26 & Calar Alto  & 3.5m      & TWIN       & C. Karl, T.
Lisker \\
2002-06-04 to 05 & Paranal     & UT2       & UVES       & R. Napiwotzki
\\
2002-08-11 to 16 & Calar Alto  & 3.5m      & TWIN       & C. Karl \\
2002-09-23 to 26 & La Silla    & NTT       & EMMI       & C. Karl \\
2002-10-15 to 16 & Calar Alto  & 3.5m      & TWIN       & C. Karl \\
2002-11-10 to 13 & Paranal     & UT2       & UVES       & R. Napiwotzki
\\
2003-01-22 to 25 & La Palma & WHT & ISIS & G. Nelemans, P. Groot\\
2003-02-17 to 21 & La Silla    & NTT       & EMMI       & C. Karl \\
2003-03-17 to 20 & Paranal     & UT2       & UVES       & R. Napiwotzki
\\
2003-03-20 to 23 & La Palma & WHT & ISIS & G. Nelemans, R. Izzard \\
2003-04-14 to 20 & Calar Alto  & 3.5m      & TWIN       & C. Karl \\
2003-05-13 to 18 & Calar Alto  & 3.5m      & TWIN       & C. Karl \\
2003-07-01 to 17 & Calar Alto  & 3.5m      & TWIN       & C. Karl \\
2003-08-06 to 09 & La Palma & WHT & ISIS & G. Nelemans, M. Montgomery \\
2004-01-01 to 04 & La Silla    & NTT       & EMMI       & C. Karl \\
2004-11-27 to 01 & La Silla    & NTT       & EMMI       & C. Karl \\
2004-12-03 to 05 & La Palma    & WHT       & ISIS       & G. Roelofs \\
2004-12-25 to 30 & La Palma    & WHT       & ISIS       & G. Roelofs \\\hline
\end{tabular}
\end{center}
\end{table*}

Most binaries in which the two components are close enough together to
interact at some stage, will end their life as close double white
dwarfs \citep[e.g.][]{web84}. They will experience at least two phases
of mass transfer.  In at least one of these phases, a so-called common
envelope will be formed, in which the first formed white dwarf
spirals into the envelope of its companion \citep[e.g.][]{nyp+00}.
  
Although this population of close double white dwarfs was expected on
theoretical grounds \citep{web84}, it was only in 1988 that the first
system was discovered \citep{slo88}. In the 1990s the number of known
systems increased to 18 \citep[see][]{mar00,mmm00}.  Detailed
knowledge of this population is important for a number of reasons.
The observed double white dwarfs carry important information about the
preceding binary evolution; in particular the mass ratio of the
components reflects the change in orbital separation in the mass
transfer phases and puts constraints on the physics and applicability
of the common envelope phase \citep{nvy+00}.  Furthermore double white
dwarfs are so abundant that they form an important background for
gravitational wave radiation experiments like LISA, as well as being
guaranteed detectable sources \citep[e.g.][]{nyp03}. Last but not
least double white dwarfs play an important role in the still ongoing
debate on, or rather quest for, the nature of the progenitors of type
Ia supernovae.
  
Type Ia supernovae (SNe Ia) are among the best standard candles to
determine the cosmological parameters \citep[see][]{lei01}.  Yet the
nature of the SNe~Ia progenitors remains unclear \citep[see][for a
review]{liv01}. One of the progenitor models is the merger of two
white dwarfs \citep{it84b}.  Some explosion calculations show that
merging double white dwarfs do \emph{not} lead to an explosive event
\citep[e.g.][]{sn98} even when rotation is taken into account
\citep{sn04}, while other calculations including rotation do get
explosions \citep{pgi+03}.  The current explosion models are still too
simplistic to rule out double white dwarfs as progenitors. For the use
of SNe~Ia as standard candles the nature of the progenitors is crucial
\citep[e.g.][]{hwt98}. One basic question that needs to be answered
for any possible progenitor scenario is whether there are enough of them.
  
To give a definite answer to the question whether there exist enough
merging double white dwarf with a mass above the Chandrasekhar limit
we embarked on a spectroscopic survey of white dwarfs with the UVES
spectrograph at the ESO-VLT \citep[ESO SNIa Progenitor Survey,
][]{ncd+01} and have observed more than 1000 white dwarfs (which
corresponds to about 75 per cent of the known white dwarfs accessible
by the VLT with magnitudes brighter than B = 16.5). For each object we
took two spectra to check for binarity, in order to provide good
statistics of the double white dwarf population, including SN Ia
progenitor candidates. A total of 875 stars were checked for radial
velocity variations. SPY detected $\approx100$ new double white dwarfs
\citep[see][]{nyn+04}. There are 16 double-lined systems and 18 non-DA
white dwarf binaries. The sample furthermore turned out to be
contaminated with some sdB stars, 31 of which are binaries
\citep{nkl+04}. The SPY project obtains 2 to 3 spectra per object,
enough to determine its binarity and pick out the 1 in 10 or so white
dwarfs that reside in a close binary. However, in order to obtain
orbital solutions of all the binaries discovered in the SPY project,
we started a systematic study of all new binaries, mainly with smaller
telescopes. Earlier results were published in
\citet{neh+01,nkn02,knn03}. An overview of the follow-up observations
up to Dec 2004 is given in table~\ref{tab:obs}. In this paper we
report results for the first five single-lined double white dwarfs for
which we found the orbital solutions. Results on other types of
binaries (double-lined systems and sdB binaries) will be presented in
separate papers. In Sect.~\ref{method} we discuss our observations,
data reduction and analysis, in Sect.~\ref{results} we present the
orbital solutions for the five systems and in Sect.~\ref{discussion}
we \rev{briefly} compare our findings with population synthesis
calculations.

\section{Method}\label{method}

\subsection{Observations and data reduction}

Data were taken with a number of different telescope/instrument
combinations (see Table~\ref{tab:obs}). The details of the different
setups are as follows.

\vspace*{-0.3cm}
\paragraph{UT2/UVES:} 
data taken with the UVES echelle spectrograph on UT2 of the 8.2m Very
Large Telescope of the European Southern Observatory at Paranal with
the following setup: dichroic 1, central wavelengths 3900 and 5640\AA\
giving an almost full spectral coverage from 3200 to 6650\AA\ with two
small ($\sim80$\AA) gaps around 4580 and 5640\AA. Slit widths varied
between 1" and 1.5" for the follow-up observations and 2.1" for the
survey spectra, yielding a spectral resolution of 0.36\AA\ at
H$\alpha$ or better \citep[see][]{ncd+01}.

\vspace*{-0.3cm}
\paragraph{3.5m/TWIN:} data taken with the double arm TWIN spectrograph on
the 3.5m telescope at Calar Alto. Only data from the red arm are used
in this paper. The setup consisted of the 230 mm camera with the T06
grating (1200 grooves mm$^{-1}$) and a SITe-CCD giving a dispersion of
0.55 \AA/pix.

\vspace*{-0.3cm}
\paragraph{INT/IDS:} 
data taken with the Intermediate Dispersion Spectrograph in the 2.5m
Isaac Newton Telescope on La Palma. The 500 mm camera with the 1200R
grating centred on H$\alpha$ was used, giving a dispersion of
0.39\AA/pix.

\vspace*{-0.3cm}
\paragraph{WHT/ISIS:} 
data taken with the double arm ISIS spectrograph on the 4.2m William
Hershel Telescope on La Palma. Only data from the red arm are used in
this paper. The setup was the 1200R grating centred on H$\alpha$,
together with the MARCONI CCD giving a dispersion of 0.23 \AA/pix.

\vspace*{-0.3cm}
\paragraph{NTT/EMMI:}
 data taken with the EMMI spectrograph on the New Technology Telescope
at La Silla in the red medium dispersion mode (REMD). We used grating
\#6, giving a dispersion of 0.2\,\AA/pix on the MLT/LL CCD.

Data reduction was done using standard IRAF or MIDAS tasks. Bias
subtraction was done using the overscan region, while flatfielding was
done using daytime dome flats. Spectra were extracted using optimal
extraction, except for the survey spectra. Wavelength calibration is
based on CuArNe arcs taken at the same position as the object before
or after each exposure (for the IDS/ISIS data) and ArTh for the TWIN
data. Details of the UVES data reduction will be described in a
forthcoming paper on the SPY radial velocity catalogue (Napiwotzki et
al, in prep).

\subsection{Radial velocity measurement and period determination}

Based on previous experience we have used two different techniques of
measuring the radial velocities of the white dwarfs.

The first is described in \citet{nkn02} and uses functions within the
MIDAS context to fit a combination of a Gaussian plus a Lorentzian to
the spectra. The radial velocities are calculated from the central
wavelength of the fitted Gaussians. The velocities and observation
times were later corrected to heliocentric values. In order to
determine the orbital parameters radial velocity curves for a range of
periods are fitted to produce a ``power spectrum'' indicating the fit
quality as function of period.

We also used the technique developed by Marsh and collaborators, which
utilizes a multi-Gaussian fitting technique to fit the shape of the
spectral lines \citep[see][]{mdd95}. A fixed template consisting of a
number of different Gaussians (typically two or three), is fitted to
the individual spectra in order to measure the radial velocity. A
floating mean period search is used to find the orbital solution by
minimizing the $\chi^2$ of the fit of a sinusoidal curve to the radial
velocity measurements \citep[see][for a discussion of this
technique]{mmm+04}. This yields the best fit period, radial velocity
amplitude, velocity offset (in principle the system velocity, or
$\gamma$-velocity, but not corrected for gravitational redshift), the
zero-point (where the star moves from the blue to the red) and the
mass function.

In order to cross check the two analysis methods we compared the
resulting radial velocity measurements in detail for two binaries
(HS1102+0934 and HS1102+0032, details of which will be published
later) and found good agreement. Most values agree within 1 sigma, the
maximum deviation was 1.44 sigma. All radial velocities presented in
this paper are obtained with the second method.

Before discussing the individual objects we spend a few words on the
error analysis. Following earlier practice with radial velocity data
of double white dwarfs, we assume there are systematic errors of a few
km s$^{-1}$ that are due to imperfections in the setup. We therefore
added 2 km s$^{-1}$ quadratically to the errors that we obtain by
propagating the errors on the spectra. We found that there were some
additional errors that arise from combining data from different
telescopes/instruments. We did not include any additional error term
to account for this, but we chose to keep the errors as they are,
which in some cases results in rather large values of $\chi^2$. The
errors on the derived quantities might thus be a bit
optimistic. However, none of the derived periods is uncertain due to
this effect.


\begin{table*}
\caption{Overview of results of the follow-up observations. The
  columns give orbital period ($P_{\rm orb}$), the radial velocity
  semi-amplitude ($K$), the $gamma$-velocity ($\gamma$), the
  zero-point ($T_0$, where the star moves from the blue to the red),
  the $\chi^2$ of the fit, the number of data points (N), the
  difference of the $\chi^2$ with the $\chi^2$ of the second best fit
  period ($\Delta \chi^2$), the mass of the visible white dwarf
  ($M_1$; from Voss et al. in prep.) and the minimum mass of the
  companion white dwarf, from the mass function, $M_1$ and assuming an
  inclination of 90 degrees ($M_{\rm 2, min}$).  }
\begin{center}
\begin{tabular}{ldmmddrrdd}\hline \hline
Object       &\multicolumn{1}{r}{$P_{\rm orb}$}& \multicolumn{1}{r}{$K$}           & \multicolumn{1}{r}{$\gamma$}     & \multicolumn{1}{r}{HJD (T$_0$)} & \multicolumn{1}{r}{$\chi^2$} & N  & $\Delta \chi^2$ & \multicolumn{1}{r}{$M_1$} & \multicolumn{1}{r}{$M_{2,{\rm min}}$} \\
             &\multicolumn{1}{r}{ (d)}         &\multicolumn{1}{r}{km s$^{-1}$} & \multicolumn{1}{r}{km s$^{-1}$}  & \multicolumn{1}{r}{$-$2400000}    &          &    &                 & \multicolumn{1}{r}{\msun} & \multicolumn{1}{r}{\msun}             \\  \hline
HE0320$-$1917  & 0.86492(4)  & 105+1   & 43+1     & 52658.97(5) & 33.0     &15  & 290             & 0.29 & 0.35              \\
HE1511$-$0448  & 3.222(0)    & 74+4    & -36+2  & 52270.1(6)  & 12.9     &15  & 31              & 0.48 & 0.46              \\
WD0326$-$273   & 1.8754(5)   & 96.2+0.5& 70.0+0.4 & 52165.23(9) & 69.8     &39  & 1000         & 0.51  & 0.59              \\
WD1013$-$010   & 0.43653(5)  & 122+2   & 59+1     & 53016.41(1) & 22.3     &15  & 170              & 0.44 & 0.38              \\
WD1210+140   & 0.64194(3)  & 131+3   & 15+2     & 52170.29(4) & 26.3     &15  & 50               & 0.23 & 0.38              \\ \hline
\end{tabular}
\end{center}
\label{tab:results}
\end{table*}

\begin{figure}
\centering
\includegraphics[angle=-90,width=\columnwidth]{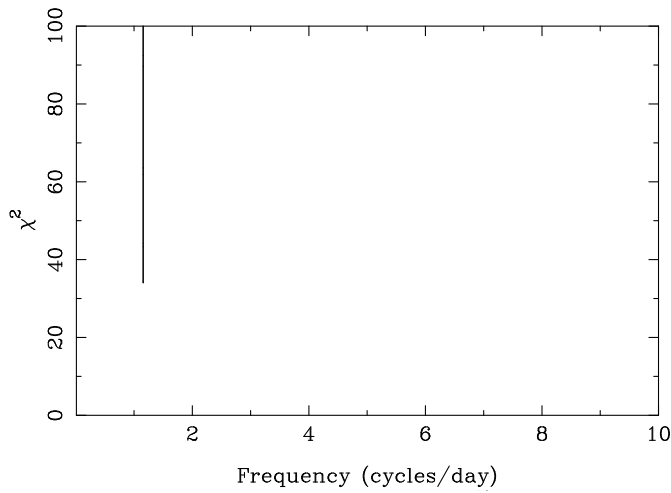}
\includegraphics[angle=-90,width=\columnwidth,clip]{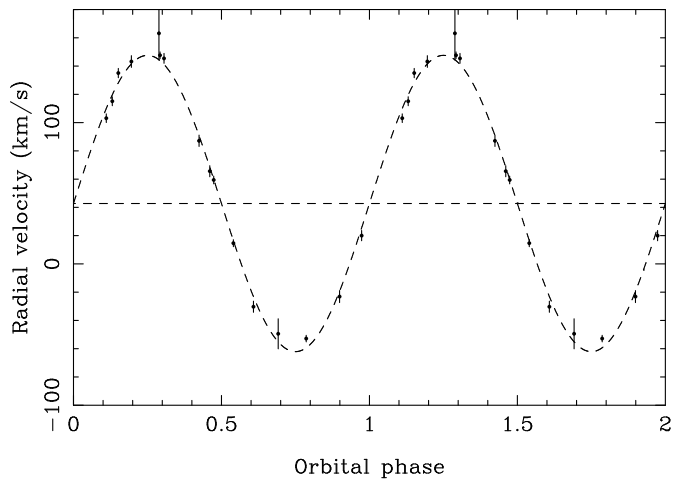}
\caption{{\it Top:} periodogram for HE0320$-$1917 showing the $\chi^2$ as a
function of the orbital period in cycles per day. Due to the
availability of measurements over a long time span the $\chi^2$ minima
are so narrow that they appear as vertical lines. {\it Bottom:} the Radial
velocity data folded on the best fit period.}
\label{fig:he0320}
\end{figure}

\section{Results}\label{results}

Table~\ref{tab:results} summarizes the results for this first set of
double white dwarfs. Listed are the orbital period ($P_{\rm orb}$),
the radial velocity semi-amplitude ($K$), the system-velocity
($\gamma$), the zero-point ($T_0$, where the star moves from the blue
to the red), the $\chi^2$ of the fit, the number of data points (N),
the difference of the $\chi^2$ with the $\chi^2$ of the second best
fit period ($\Delta \chi^2$), the mass of the visible white dwarf
($M_1$) and the minimum mass of the companion white dwarf, from the
mass function, $M_1$ and assuming an inclination of 90 degrees
($M_{\rm 2, min}$). 

The masses $M_1$ of the visible component were determined from a model
atmosphere analysis of th UVES survey spectra (Voss et al.\ 2005, in
prep.). In short, we determined temperature and surface gravity by
fitting model spectra to the observed hydrogen Balmer lines of the DA
white dwarfs.  Depending on the temperature of the star we applied LTE
or NLTE model atmospheres \citep[cf.][for a
description]{knc+01}. Finally we interpolated in the white dwarf
cooling tracks of \citet{ba99} to determine the mass. The results for
the systems are discussed below individually.

\begin{figure}
\centering
\includegraphics[angle=-90,width=\columnwidth]{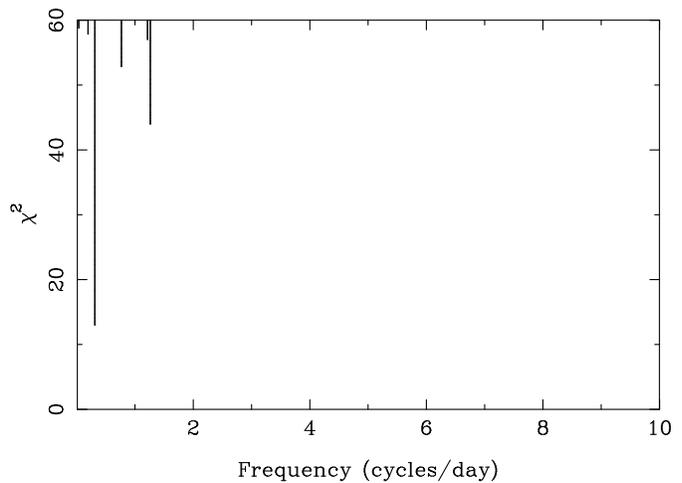}
\includegraphics[angle=-90,width=\columnwidth,clip]{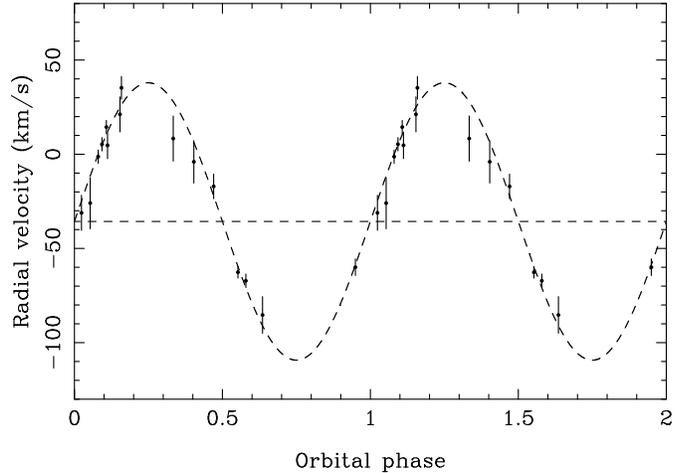}
\caption{Periodogram for HE1511$-$0448 in cycles per day (top) and data folded on the best fit period.}
\label{fig:he1511}
\end{figure}

\subsection{HE0320$-$1917}

HE0320$-$1917 is relatively cool white dwarf of about 12,000 K, with a
mass of 0.29 \msun. It was selected from the Hamburg/ESO survey
\citep[see][]{cwr+01}. The periodogram of HE0320$-$1917 is shown in
Fig.~\ref{fig:he0320}. The best fit period is about 0.865 d (see
Table~\ref{tab:results}).  The radial velocity measurements folded on
the best fit period are shown in Fig.~\ref{fig:he0320}. The minimum
mass of the unseen companion is 0.35 \msun.

\subsection{HE1511$-$0448}

HE1511$-$0448 was again selected from the Hamburg/ESO survey and is a
very hot white dwarfs (50,000 K), with a mass of 0.48 \msun.  The line
profile shows an emission core, so this object was fitted with three
Gaussians, one for the emission core. The periodogram of HE1511$-$0448
is shown in Fig.~\ref{fig:he1511}. The orbital period is 3.22
days. There are two other solutions for the period (0.8 and 1.3 d),
but they are very unlikely. The folded radial velocity curve is shown
in Fig.~\ref{fig:he1511}.


\subsection{WD0326$-$273}
\label{wd0326}

\begin{figure}
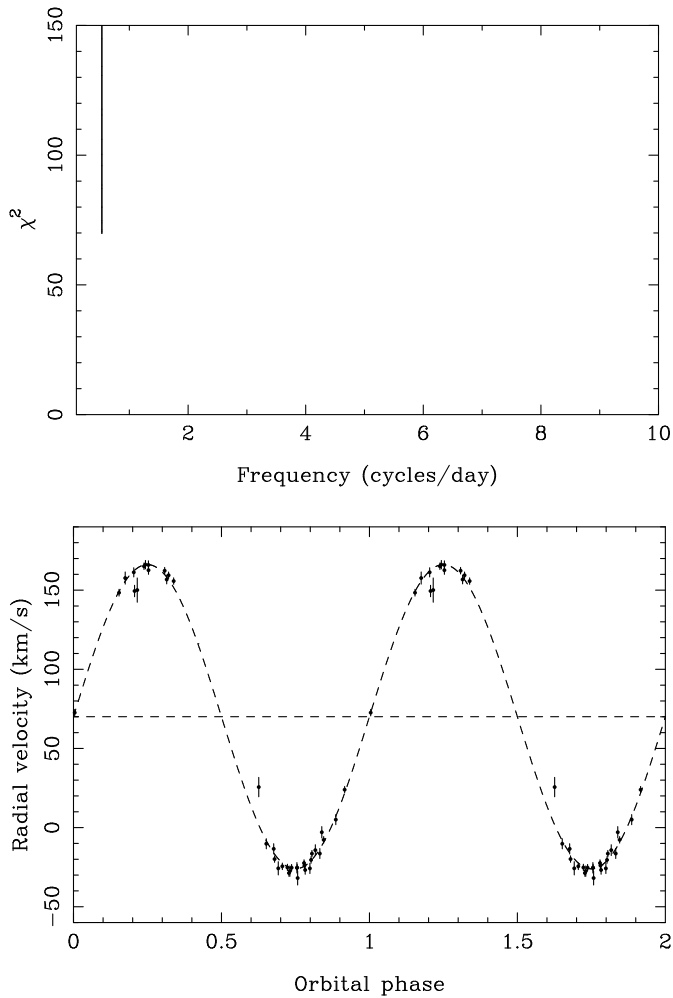

\centering
\includegraphics[angle=-90,width=\columnwidth]{3174f5.ps}
\includegraphics[angle=-90,width=\columnwidth,clip]{3174f6.ps}
\caption{Periodogram for WD0326$-$273 in cycles per day (top) and data
folded on the best fit period.}
\label{fig:wd0326}
\end{figure}

WD0326$-$273 (GJ 1060A, L\,587$-$77A) was discovered by Luyten as a
high proper motion white dwarf with an M(3.5) companion at $\sim$7
arcsec \citep[e.g.][]{luy49}. The velocity difference of the two stars
has been used to determine the gravitational redshift of the white
dwarf \citep{koe87}, yielding a mass of 0.65 \msun. \citet{wr91} on
the other hand measured a blue shift and discarded that system as not
physically associated. However, both these results are, of course,
influenced by the orbital velocity.

\begin{figure}
\centering
\includegraphics[angle=0,width=\columnwidth,clip]{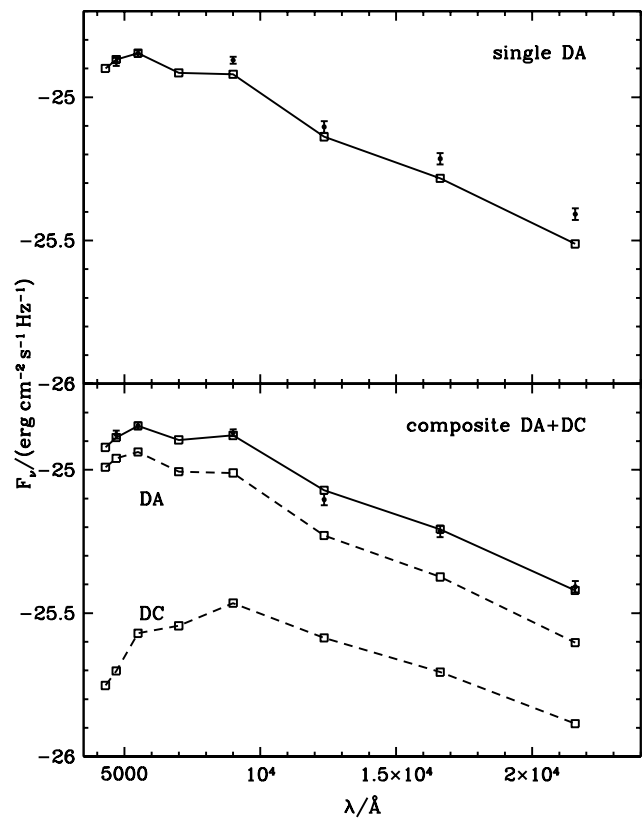}
\caption{{\it Top:} Model flux of a 9300\,K DA white dwarf compared to
the observed fluxes (converted from magnitudes). For clarity we show
only the JHK measurements of \citet{blr01}. {\it Bottom:} Comparison
of the best fit composite DA+DC spectrum to the observed fluxes. The
individual contributions of the DA and DC components are shown as
dashed lines.}
\label{fig:wd0326flux}
\end{figure}

Spectral fits to the SPY UVES spectra yield a mass of $0.46 M_\odot$
(Voss et al.\ 2005, in prep.).  \citet{blr01} find a mass of 0.35
\msun based on photometric and parallax data, but a higher mass using
spectroscopic mass determination. The analysis of the photometric data
was flawed, because Bergeron et al.\ adopted $V=14.0$, a value which
can be traced back to a photographic measurement by \citet{luy57}. A
more accurate value \rev{(y/V=13.56)} is available from photoelectric
observations in the Str\"omgren system by \citet{bw78}. We will
discuss evidence for a remaining infrared excess below.

\vspace*{-0.3cm}
\paragraph{Gravitational redshift:} With our orbital solution we are
now in a position to determine the real gravitational redshift of the
visible DA white dwarf. Due to a (fortunate) target confusion during
the SPY survey observations one UVES spectrum of the M star companion
L\,587$-$77B was taken. The Balmer lines are in emission and we
measured the radial velocity from a simple Gaussian fits of the
H$\alpha$ and H$\beta$ emissions. The measured value $v_{\rm M
star} = 47\pm 1$\,km\,s$^{-1}$ translates into a redshift of
23\,km\,s$^{-1}$. Applying the white dwarf cooling tracks of
\citet{ba99} we derive a white mass of $0.51M_\odot$, slightly higher
than the spectroscopic result. Let us note that taken at face value
our radial velocity determination of the M dwarf is discrepant to the
\citet{wr91} measurement of $v_{\rm M star} = 17\pm
10$\,km\,s$^{-1}$. However, the authors caution that the errors of
their absolute radial velocity measurements are probably larger than
the purely instrumental error.

\begin{figure}
\centering
\includegraphics[angle=-90,width=\columnwidth]{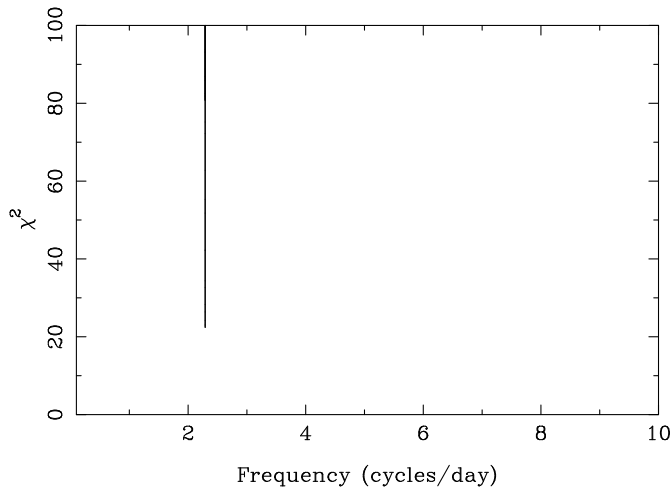}
\includegraphics[angle=-90,width=\columnwidth,clip]{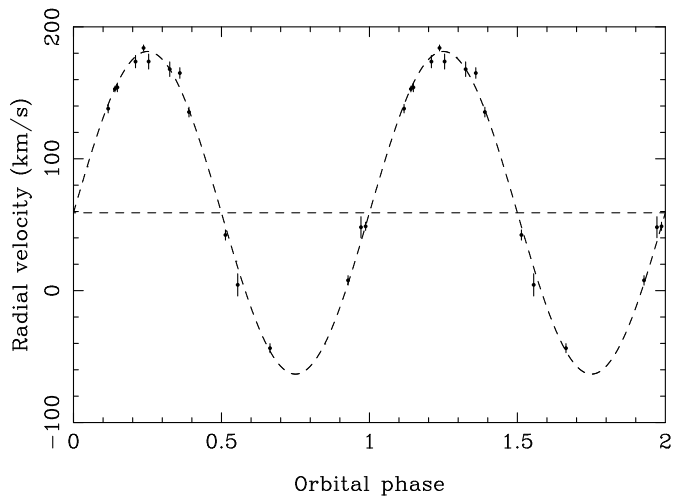}
\caption{Periodogram for WD1013$-$010 in cycles per day and data folded on the best fit period.}
\label{fig:wd1013}
\end{figure}

\begin{figure}
\centering
\includegraphics[angle=-90,width=\columnwidth]{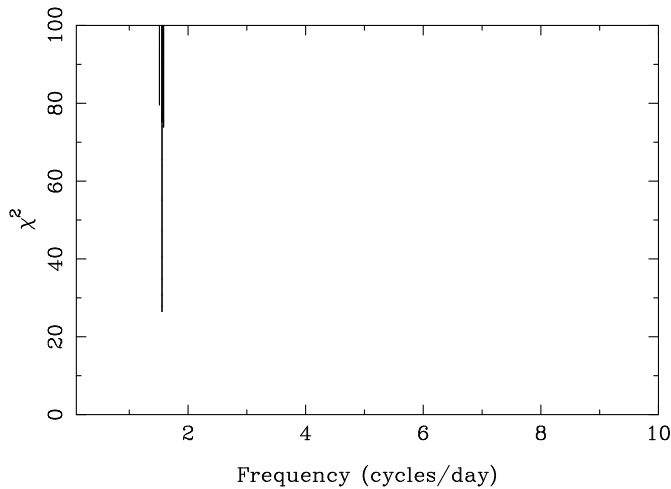}
\includegraphics[angle=-90,width=\columnwidth,clip]{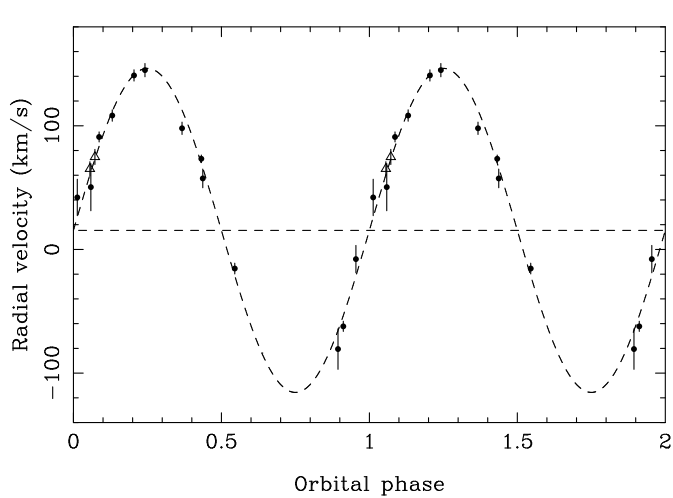}
\caption{Periodogram for WD1210+140 in cycles per day and data folded on the best fit period.}
\label{fig:wd1210}
\end{figure}

\vspace*{-0.3cm}
\paragraph{The cool white dwarf companion:} The quality of the model
atmosphere fit of WD\,0326$-$273 is inferior to the quality typically
achievable for white dwarfs in this parameter range ($T_{\mathrm{eff}}
= 9300$K, Voss et al.\ 2005, in prep.). Closer inspection of the
optical and IR photometry revealed that a significant red/infrared
excess is present (Fig.~\ref{fig:wd0326flux}, top).  \rev{A plausible}
explanation is a flux contribution from the apparently invisible cool
white dwarf companion of the brighter DA.  We collected the available
photometric measurements (Str\"omgren $b$ and $y$ from \citet{bw78};
$J,H,K$ from \citet{blr01}; $I$ and $J$ from DENIS; $J,H,K$ from
2MASS) and tried to fit these with a composite spectrum. We fixed the
temperature of the bright DA component at the spectroscopic value of
9,300\,K and adopted equal masses for both white dwarfs. Synthetic
absolute magnitudes and colours for pure hydrogen and pure helium
atmospheres were taken from \citet{bwb95}. Fluxes were co-added and
normalised to the measured V flux. A $\chi^2$ analysis yielded a best
fit for a helium-composed companion with a temperature
$T_{\mathrm{eff}} = 6150\pm 650$\,K (displayed in the bottom panel of
Fig.~\ref{fig:wd0326flux}).  This corresponds to a featureless DC
white dwarf. Fitting the colours for the hydrogen-rich models yielded
a slightly worse fit at $T_{\mathrm{eff}} = 7300\,\pm 350$\,K. However,
this solution can be ruled out, because it predicts a relatively
strong H$\alpha$ line for the secondary, which is clearly not present
in the observed spectra (see Fig.~\ref{fig:wd0326_quad}).

The inner binary (assuming the M3.5 star is physically associated in a
very long period outer binary) has an orbital period of 1.88 d
(Fig.~\ref{fig:wd0326}). There are no other periods possible, given
the relatively large number of spectra we have of this source. The
large amplitude of the radial velocity curve shows that the companion
star also has to be rather massive (more massive than 0.59 \msun, see
Table~\ref{tab:results}).

\subsection{WD1013$-$010}

WD1013$-$010 is a cool white dwarf (8,000 K) with a mass of 0.44
\msun. The periodogram for WD1013$-$010 is plotted in
Fig.~\ref{fig:wd1013}. The best fit period is 0.437 d. The folded
radial velocity curve is shown in Fig.~\ref{fig:wd1013}. The next best
period, with $\chi^2$ = 51 has virtually the same period and thus
overlaps with the best fit period in Fig.~\ref{fig:wd1013}.

\subsection{WD1210+140}

WD1210+140 is another hot white dwarf (32,000 K) of very low mass:
0.23 \msun. It is the lowest mass white dwarf known in a double white
dwarf binary, and one of the lowest mass white dwarfs known.
Interestingly, \citet{mm99} observed WD1210+140 and, despite the large
radial velocity amplitude of the system, concluded is is not a binary,
but only based on two (consecutive) spectra. We included their two
measurements (taken in 1998) in our fit. This results in a unique
solution for the system, with orbital period of 0.64 d.
(Table~\ref{tab:results}). The radial velocity curve, folded on the
best period is shown in Fig.~\ref{fig:wd1210}. \rev{The measurements
  of \citet{mm99} are included as open triangles. The mass of
  WD1210+140 is virtually identical to the mass of the sdB star in the
  0.6 d binary HD 188112, in which the companion is a massive ($M >
  0.73 \msun$) compact object \citep{hel+03}.}

\section{Is there any sign of the companions?}

\begin{figure}
\centering
\includegraphics[angle=-90,width=\columnwidth,clip]{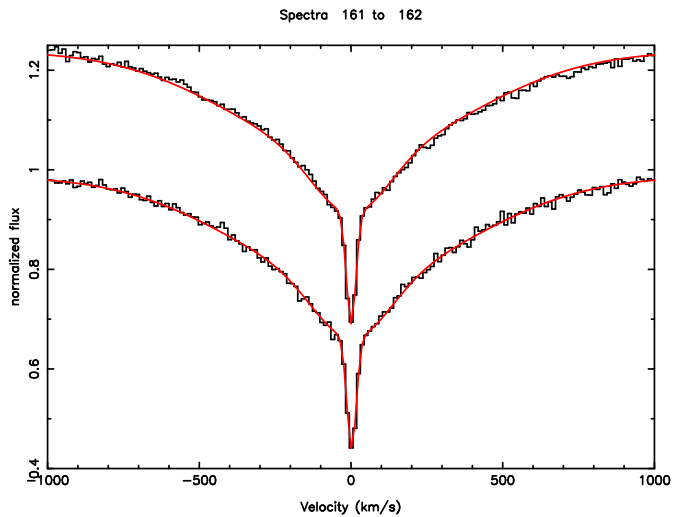}
\caption{Quadrature spectra of WD0326$-$273, after shifting out the best fit
sinusoidal radial velocity curve and combining the spectra in the
phase intervals 0.15--0.35 (top) and 0.65--0.85 (bottom). Any signs of
the companion would show up as an asymmetry in these two spectra.}
\label{fig:wd0326_quad}
\end{figure}

In order to search for signs of weak absortion lines of the
companions, that potentially could be used to determine the mass
ratio, we have to inspect the spectra around quadrature (phases 0.25
and 0.75) when the two spectra are most widely separated. In order to
obtain high quality spectra to look for weak spectral features one
generally combines spectra with similar phases.  However, in general
our single high singnal-to-noise ratio VLT survey spectra, that were
used to measure the initial radial velocity offset are more suitable
for this. In the survey spectra none of the systems discussed here
showed any sign of absorption lines of the companion.  Only for
WD0326$-$273 we can improve on these results as many of the follow-up
spectra were taken with the VLT as well. We therefore combined the VLT
spectra in the phase range 0.15--0.35 (4 spectra) and 0.65--0.85 (9
spectra) after shifting out the radial velocities according to our
best fit orbital solution. The resulting spectra are shown in
Fig.~\ref{fig:wd0326_quad}, no additional absorption lines are seen,
but see Sect.~\ref{wd0326} for our arguments that there must be
significant flux from a featureless companion.

\section{Discussion}\label{discussion}

\begin{table}
\caption[]{Properties of the observed double white dwarfs}
\label{tab:dwd}
\begin{center}
\begin{tabular}{lddrr} \hline
Object   & \multicolumn{1}{c}{$P$} &
\multicolumn{1}{c}{$M_{1}$} & \multicolumn{1}{c}{$M_{\rm 2}$} &  Ref \\ 
 &  \multicolumn{1}{c}{(d)} &
\multicolumn{1}{c}{(\msun)} & \multicolumn{1}{c}{(\msun)} & \\ 
\hline
WD0135$-$052 &  1.56 &  0.47 &   0.52 & 1,2  \\
WD0136$+$768 &  1.41 &  0.47 &   0.37 & 8,13 \\
HE0320$-$1917&  0.87 &  0.29 &$>$0.35 & 18 \\
WD0326$-$273 &  1.88 &  0.51 &$>$0.59 & 18 \\
WD0957$-$666 &  0.06 &  0.37 &   0.32 & 3,7  \\
WD1013$-$010 &  0.43 &  0.44 &$>$0.38 & 18 \\
WD1022$+$050 &  1.16 &  0.39 &$>$0.28 & 8,17 \\
WD1101$+$364 &  0.15 &  0.29 &   0.35 & 4,13 \\
WD1115$+$166 & 30.09 &  0.52 &   0.43 & 12   \\
WD1202$+$608 &  1.49 &   0.4 &$>$0.34 & 6    \\
WD1204$+$450 &  1.60 &  0.46 &   0.52 & 8,13 \\
WD1210$+$140 &  0.64 &  0.23 &$>$0.38 & 18  \\
WD1241$-$010 &  3.35 &  0.31 &$>$0.37 & 5    \\
WD1317$+$453 &  4.87 &  0.33 &$>$0.42 & 5    \\
WD1349$+$144 &  2.12 &  0.44 &   0.44 & 14   \\
HE1414$-$0848&  0.52 &  0.71 &   0.55 & 11,15\\
WD1428$+$373 &  1.14 &  0.35 &$>$0.23 & 9,17 \\
HE1511$-$0448&  3.22 &  0.48 &$>$0.46 & 18 \\
WD1704$+$481 &  0.14 &  0.39 &   0.56 & 10   \\
WD1713$+$332 &  1.12 &  0.35 &$>$0.18 & 5    \\
WD1824$+$040 &  6.27 &  0.43 &$>$0.52 & 8,17 \\
WD2032$+$188 &  5.08 &  0.41 &$>$0.47 & 8,17 \\
HE2209$-$1444&  0.28 &  0.58 &   0.58 & 16   \\
WD2331$+$290 &  0.17 &  0.39 &$>$0.32 & 5    \\ \hline
\end{tabular}
\end{center}

References: (1) \citet*{slo88}; (2) \citet{bwf+89}; (3) \citet{bgr+90};
(4) \citet{mar95}; (5) \citet*{mdd95}; (6) \citet{hst+95}; (7)
\citet*{mmb97}; (8) \citet{mm99}; (9) \citet{mar00} and P. Maxted,
private communication; (10)
\citet{mmm+00}; (11) \citet{nkn02}; (12) \citet{mbm+02}; (13)
\citet{mmm02}; (14) \citet{knh+02}; (15) \citet{ndh+02}; (16)
\citet{knn03}; (17)\citet{mmm+04}; (18) this paper
\end{table}

In total there are now 24 double white dwarfs known. The parameters of
all systems are shown in Table~\ref{tab:dwd}, including the new
systems discussed in this paper.  Of the new systems WD0326$-$273 and
WD1210+140 are particularly interesting, having rather high (0.51
\msun) and very low (0.23 \msun) masses respectively. The difference
in the approach between the earlier work (targeting low-mass white
dwarfs) and the SPY project is starting to become clear. Of the four
clear C/O white dwarfs (taken as $M > 0.5 \msun$) three were found by
SPY. There are a few changes compared to earlier compilations like
this \citep[e.g.][]{mm99,nt03}, in particular some of the masses of
the white dwarfs have been redetermined.  There are also some small
changes in the masses given in new papers \citep{mmm02,mmm+04}. In the
table we also list the minimum companion star mass for the
single-lined spectroscopic binaries, based on the mass function and
assuming an inclination of 90 degrees. We can now use the above data
in order to compare with theoretical models of the local population of
double white dwarfs \rev{to see if the new systems change the
conclusions of earlier comparisons}. We compare the data with a
slightly updated version of the population synthesis calculation of
\citet{nyp+00}, as detailed in \citet{nyp03}. \rev{Earlier population
synthesis calculations for double white dwarfs are reported e.g. in
\citet{ity97,han98}.}

\begin{figure}
\centering
\includegraphics[angle=-90,width=\columnwidth]{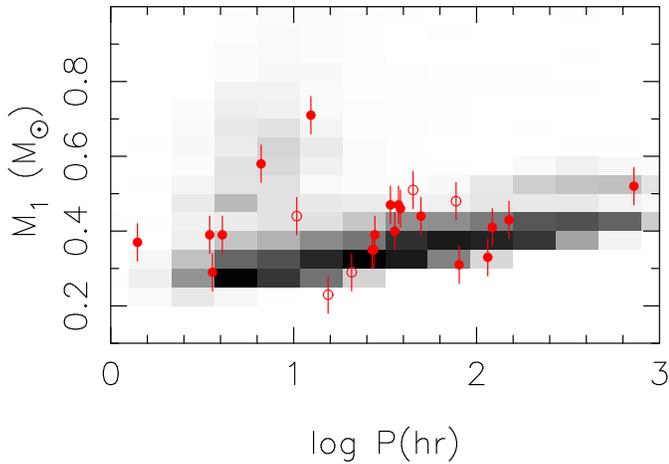}
\caption{Period - mass \rev{($M_1$)} distribution for double white
  dwarfs.  The systems presented in this paper are shown with the open
  circles.}
\label{fig:Pm}
\end{figure}

In Fig.~\ref{fig:Pm} we plot the distribution in the period - white
dwarf mass plane, for the brighter white dwarf in each binary
($M_1$). All observed double white dwarfs are plotted on top of the
theoretical distribution for a magnitude limited sample with ($V <
20$). A few things should be noted before conclusions based on this
diagram can be drawn. The most important is that the observed systems
are not (yet) an unbiased sample. The SPY project will provide a much
more unbiased sample, but only when the majority of the discovered
binaries have known periods. More specific, the observed systems are
still dominated by the low-mass white dwarfs that were targeted by
Marsh and collaborators.

Taking the plot as it is, two things can be concluded. The first is
that the new observations \rev{occupy similar regions in the diagram
and thus} \2rev{conform to} the general picture that was seen before
and which can be explained adequately by the population synthesis
calculations.  It also confirms the earlier finding \citep{nyp+00}
that the lowest mass white dwarfs do not keep a thick hydrogen
envelope in which residual burning continues, as was suggested by
\citet{dsb+98}, which would predict many systems with very low masses
\rev{(although WD1210+140 shows that some systems with $M < 0.25
\msun$ do exist)}. This might be due to thermal instabilities in which
the hydrogen is lost \citep[e.g.][]{dbs+99}. However, we currently use
a cooling prescription based on the models of \citet{han99}, as
detailed in \citet{nyp03}, which assume a constant hydrogen envelope
mass as function of the mass of the white dwarf. This prescription
enhances the presence of systems with masses above 0.5 \msun in
Fig.~\ref{fig:Pm}, as compared to the same plot in \citet{nyp+00},
more consistent with the observations. The second conclusion \rev{is
that the apparent gap in the period distribution between 0.5 and 1
day, which was noted in \citet{nyp+00} indeed was due to small number
statistics and is filling in.}

\begin{figure}
\centering
\includegraphics[angle=-90,width=\columnwidth]{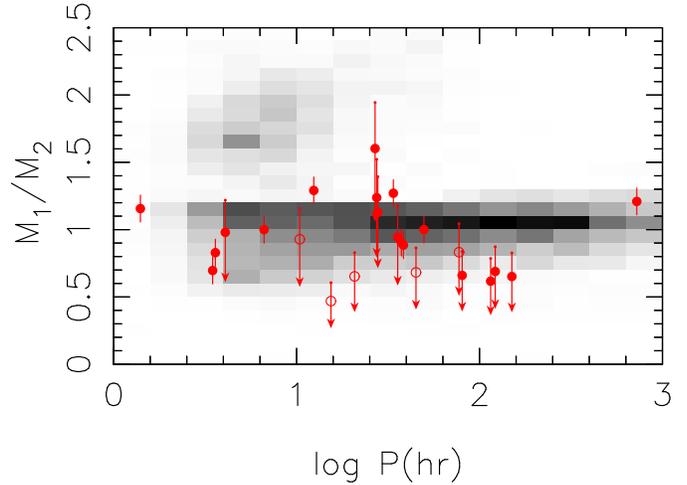}
\caption{Period - mass ratio \rev{($M_1/M_2$)} distribution for double
  white dwarfs.  The systems presented in this paper are again shown
  with the open circles. The systems for which only a lower limit to
  the mass of the companion is know are shown with the arrows. They
  are drawn from the maximum mass ratio (for $M_2 = M_{\rm 2,min}$),
  to the point where there is a 75 per cent probability that the mass
  ratio lies above this point. The large circle is plotted at the
  position of the median inclination (60 degrees). Open circles again
  are the systems presented in this paper. \rev{In the theoretical
  distribution $M_1$ is selected to be the mass of the brightest
  object of the two (based on age and cooling, generally the second
  formed white dwarf).} }
\label{fig:Pq}
\end{figure}

In Table~\ref{tab:dwd} we also included the minimum mass of the
companion for the single-lined systems. The mass ratios of double
white dwarfs\footnote{\rev{Traditionally taken to be the mass of the
brighter over the mass of the dimmer component ($M_1/M_2$),
e.g. \citet{ity97,mmm+00}}} have led to the suggestion that the
standard description of the common-envelope phase in close binaries
needs revision \citep{nvy+00}. The population synthesis calculations
of \citet{nyp+00} are based on this suggested revision. A recent
reassessment of the need for this revision, based on the newly
discovered double-lined double white dwarfs confirms the earlier
findings \citep{nt03}. In these analyses only double-lined systems
were taken into account. However, in Fig.~\ref{fig:Pq} we compare the
theoretical distribution in the period - mass ratio \rev{($M_1/M_2$)}
plane with all observations, where for the single-lined systems we
show a plausible range of mass ratios, based on the mass limit on the
companion star. For each system we plot an arrow that starts at the
maximum mass ratio (for the minimum companion star mass, i.e. for 90
degrees inclination). The arrow extends to the mass ratio for an
inclination of 41 degrees, which corresponds to a 75 per cent
probability that the actual mass ratio lies along the arrow. The large
symbol is plotted at the mass ratio for the median for random
orientations, i.e. an inclination of 60 degrees. \rev{In the
theoretical distribution for each system the brightest of the two white
dwarfs (in the V band) is determined, based on the age and the (mass
dependent) cooling curve and its mass is taken as $M_1$. This
generally is the second formed white dwarf.}

Although the general agreement between the observations and the theory
is good, it clearly is not as good for the single-lined systems as for
the double-lined systems (shown as symbols with error bars), for which
the theory was developed. In particular the mass ratios for the longer
period systems seem to be lower than predicted. \rev{Detailed
investigation of this discrepancy is beyond the scope of this paper.}

\begin{acknowledgements}
  GN is supported by NWO-VENI grant 639.041.405. C.K. was supported by
  DFG grants NA 365/3-1 and He 1356/40-3. R.N.  acknowledges support
  by a PPARC Advanced Fellowship. Calar Alto observations were
  supported by DFG travel grants. RGI thanks PPARC for a studentship,
  and CIQuA for a fellowship. GR is supported by NWO-VIDI grant
  639.042.201. B.V. acknowledges support from the Deutsche
  Forschungsgemeinschaft (grant KO738/21-1).

\end{acknowledgements}

\bibliography{journals,binaries} \bibliographystyle{aa}

\appendix
\section{Radial velocity measurements}

\begin{table}[h]
\caption{Radial velocity measurements of WD0326$-$273}
\begin{tabular}{lmclmc} \hline \hline
HJD & \multicolumn{1}{c}{RV} & Tel & HJD & \multicolumn{1}{c}{RV} & Tel \\ 
$-$2450000 & \multicolumn{1}{c}{(km s$^{-1}$)}& & $-$2450000 & \multicolumn{1}{c}{(km s$^{-1}$)}&\\\hline
1737.9246 & 148.4 +  0.8 & V  &  2213.5416 & -25.4 +  2.7 & I\\
1740.8767 & -28.7 +  0.7 & V  &  2213.6239 & -25.8 +  2.6 & I\\
2194.6455 & -19.9 +  0.5 & V  &  2540.7119 & 161.3 +  2.1 & N\\
2194.6947 & -24.5 +  0.6 & V  &  2540.7859 & 166.1 +  1.8 & N\\
2194.7268 & -25.2 +  0.5 & V  &  2540.8050 & 165.9 +  1.7 & N\\
2194.7534 & -25.3 +  0.6 & V  &  2541.7981 & -26.8 +  1.8 & N\\
2194.7886 & -25.3 +  0.5 & V  &  2541.8911 & -16.4 +  2.5 & N\\
2194.8322 & -22.4 +  0.5 & V  &  2542.6806 & 162.6 +  1.6 & N\\
2194.8820 & -16.5 +  0.4 & V  &  2542.7960 & 156.7 +  1.9 & N\\
2195.6932 & 165.1 +  0.6 & V  &  2543.6692 & -23.9 +  1.9 & N\\
2195.8242 & 162.3 +  0.5 & V  &  2543.7380 & -14.2 +  2.8 & N\\
2209.5487 &  25.6 +  5.8 & I  &  2543.7790 &  -3.0 +  3.0 & N\\
2209.5967 & -10.2 +  2.5 & I  &  2543.8676 &   5.0 +  2.5 & N\\
2209.6437 & -13.4 +  2.6 & I  &  2589.7266 & 155.7 +  0.5 & V\\
2210.5780 & 157.7 +  3.3 & I  &  2590.5973 & -20.4 +  0.7 & V\\
2210.6372 & 149.5 +  3.0 & I  &  2591.5698 & 159.4 +  0.5 & V\\
2211.5480 & -25.7 +  3.7 & I  &  2592.5529 &  -7.7 +  0.5 & V\\
2211.6212 & -27.5 +  2.6 & I  &  2592.6856 &  24.0 +  0.5 & V\\
2211.6706 & -31.9 +  3.9 & I  &  2592.8506 &  72.6 +  0.6 & V\\
2212.5297 & 150.0 +  7.4 & I  &              &              &\\
\end{tabular}
\end{table}

\begin{table*}
\caption{Radial velocity measurements of HE0320$-$1917, HE1511$-$0448, WD1013$-$010 and WD1210+140}
\begin{tabular}{lmclmclmclmc} \hline \hline
HJD & \multicolumn{1}{c}{RV} & Tel & HJD & \multicolumn{1}{c}{RV} & Tel & 
HJD & \multicolumn{1}{c}{RV} & Tel & HJD & \multicolumn{1}{c}{RV} & Tel\\ 
$-$2450000 & \multicolumn{1}{c}{(km s$^{-1}$)} & & $-$2450000 & \multicolumn{1}{c}{(km s$^{-1}$)}& & 
$-$2450000 & \multicolumn{1}{c}{(km s$^{-1}$)}& & $-$2450000 & \multicolumn{1}{c}{(km s$^{-1}$)}&\\\hline
\multicolumn{3}{l}{\textbf{HE0320$-$1917}}&   \multicolumn{3}{l}{\textbf{HE1511$-$0448}} &
\multicolumn{3}{l}{\textbf{WD1013$-$010}} & \multicolumn{3}{l}{\textbf{WD1210+140}} \\
1946.6477 &  87.1 +  3.5 &  V   & 1701.7312 & -67.2 +  3.0 & V & 2663.7956 & 184.0 + 1.1 & V &  0969.8965 &  65.2 +  5.4 &  W \\ 
1947.6129 &  14.7 +  1.4 &  V   & 2078.6244 & -62.6 +  2.4 & V & 2684.7070 & 152.9 + 0.8 & V &  0969.9073 &  74.8 +  5.8 &  W \\
2589.5967 & -52.8 +  1.2 &  V   & 2097.4761 &  -4.0 + 11.1 & C & 3337.8448 & 167.9 + 5.2 & N &  2657.8067 &  73.3 +  2.3 &  V \\
2591.7642 & 147.6 +  1.4 &  V   & 2099.4753 & -31.1 +  9.1 & C & 3337.8597 & 165.0 + 3.5 & N &  2658.7567 & -62.2 +  3.6 &  V \\
2592.6408 & 145.4 +  2.9 &  V   & 2328.6753 &  35.3 +  5.7 & C & 3338.8658 & -43.6 + 2.8 & N &  2747.4391 &  50.4 + 19.0 &  C \\
2592.7861 &  59.4 +  1.9 &  V   & 2329.6789 & -17.1 +  6.2 & C & 3339.8540 &   7.8 + 3.0 & N &  2747.5327 & 140.7 +  4.1 &  C \\
2662.3474 & -23.1 +  3.8 &  W   & 2431.5274 &  -1.3 +  2.9 & V & 3339.8730 &  48.1 + 7.7 & N &  2748.3933 & -15.5 +  3.9 &  C \\
2662.4120 &  20.0 +  3.0 &  W   & 2431.5693 &   5.3 +  2.9 & V & 3340.8233 & 154.1 + 2.6 & N &  2749.3355 &  42.1 + 14.6 &  C \\
2663.3955 & 103.1 +  2.3 &  W   & 2431.6163 &  14.4 +  3.0 & V & 3340.8695 & 173.7 + 5.3 & N &  2749.4113 & 108.3 +  4.2 &  C \\
2663.4130 & 115.1 +  2.5 &  W   & 2663.7524 &  21.2 +  9.1 & W & 3341.8019 & 135.4 + 3.1 & N &  2749.4823 & 145.0 +  5.0 &  C \\
2663.4306 & 135.0 +  2.6 &  W   & 2717.8675 & -59.9 +  3.9 & V & 3341.8558 &  42.2 + 3.3 & N &  2749.5628 &  98.0 +  4.6 &  C \\
2664.3339 & 143.2 +  3.6 &  W   & 2721.6133 &   4.7 +  6.7 & W & 3341.8739 &   4.4 + 8.2 & N &  2750.5431 & -80.5 + 16.3 &  C \\
3344.5901 & -49.4 + 10.4 &  W6  & 2825.4342 &   8.3 + 11.8 & C & 3344.7788 & 173.6 + 4.3 & W &  2750.5818 &  -7.8 + 10.9 &  C \\
3369.4729 &  65.6 +  3.4 &  W   & 2826.4060 & -85.3 +  9.5 & C & 3365.6354 &  48.8 + 2.4 & W &  2773.3597 &  57.4 +  7.3 &  C \\
3370.4654 & -30.3 +  3.4 &  W   & 2837.4181 & -26.0 + 13.4 & C & 3368.7479 & 138.0 + 2.6 & W &  3370.7838 &  91.0 +  3.5 &  W \\ \hline
\end{tabular}
\end{table*}

\end{document}